# Characteristics, Effects and Life Expectancy of a Primeval Photon

**Akinbo Ojo**
Standard Science Centre
*P.O. Box 3501, Surulere, Lagos, Nigeria*
taojo@hotmail.com

## Abstract
If a photon fluctuates from 'nothing', what would it look like and how would it behave? Here we examine such a photon and the possible characteristics and effects it may have. If contrary to theoretical expectation, it is not short-lived, we find quantitative and qualitative similarities with the quantitative and qualitative features of the standard big bang model. As it also seems devoid of some of the horizon, homogenous, flatness, singularity and temperature problems that plague the big bang model, issues surrounding this photon's life expectancy may need to be re-visited by cosmologists.

## I. Introduction
The phenomenon of a photon fluctuating from nothing is a commonly described quantum physics possibility. Speculations also abound that the universe may be a consequence of an earlier quantum fluctuation, e.g. [1,2]. However the significance of such a primeval photon for cosmology is discountenanced on the premise that from uncertainty relations, it will be short-lived, lasting only about $10^{-44}$ seconds. Leaving for the moment its lifespan, we examine the characteristics and effects of such a photon. We then follow its thermal history and make concluding inferences and remarks.

## II. Characteristics
Wavelength, $\lambda$

A photon must have a wavelength. It cannot be zero, if not it would have infinite energy, propagate at infinite velocity or not propagate at all.

If a photon fluctuates from 'nothing', i.e. a state characterized by absence of energy, matter, space and time, it would not be at liberty to pick and choose its own wavelength, at least initially. There is no pre-existing space to do this. Even if it will later vibrate at longer wavelengths, it must be content initially with the minimum, which theoretically cannot be smaller than the Planck length, $1.616 \bullet 10^{-35}$ metres.

Frequency, f

From $c = f\lambda$, where the velocity of light, $c$ is 299,792,458 $ms^{-1}$, the primeval photon must have a frequency of $1.855 \bullet 10^{43}$ Hz.

The $\bullet$ symbol stands for multiplication.

Energy, E
The initial matter-energy of this quivering photon will be $1.22 \bullet 10^{10}$ joules (~ $10^{19}$ GeV)* given that E = hf, where h is Planck's constant ($6.62 \bullet 10^{-34}$ Js).

However because it is created from nothing, it must be a closed system and its total energy must sum to zero. The positive matter-energy must therefore be balanced by the negative gravitational energy of the radius or scale factor, which will therefore be $-1.22 \bullet 10^{10}$ J in value.

There will be more on this point later, but suffice to note that it is the positive matter-energy, not the total or net energy that is responsible for the thermodynamic properties of the system, such as the temperature, ambient energy and effects on entropy.

It may also be worthy to note that since a photon cannot have a wavelength shorter than the Planck length, this is the maximum energy a single photon can have. Should the system have need to harbor higher values of matter-energy, this constraint of a shortest possible wavelength makes it inevitable that the primeval photon must be fragmented to daughter photons.

Energy density, $E_D$
Occupying a locus of the minimum possible length, if assumed spherical, this will be of Planck diameter in size. The primeval photon will thus be occupying a volume of $2.21 \bullet 10^{-105}$ $m^3$. This translates to a matter-energy density of about $5.56 \bullet 10^{114}$ $Jm^{-3}$.

Black body characteristics and temperature
Even though it has fluctuated from nothing, which state has no energy or temperature, the primeval photon is expected to have a temperature because it has positive matter-energy.

Being all there is, the primeval photon must simultaneously absorb all the radiation it emits, reflecting none. It will therefore be at equilibrium with its own radiation and we therefore expect it to behave exactly like a black body for as long as it lives. From Wien's displacement law and then the radiation density formula we estimate its temperature.

From Wien's displacement law,

$$\lambda_m T \sim 2.9 \bullet 10^{-3} \qquad (1)$$

where $\lambda_m$ is the wavelength at peak power, we obtain an estimated temperature, T of $1.79 \mathrm{X} 10^{32}$ K.

Using the radiation density formula,

$$E_D = aT^4 \qquad (2)$$

where $a$ is the radiation density constant about $7.56 \bullet 10^{-16}$ $Jm^{-3}K^{-4}$, we obtain an estimated temperature, T of $2.9 \bullet 10^{32}$ K.

*Readers should note that while some authors use this value as the empirical measure of Planck mass and energy, others use a different value which is smaller by a factor of $2\pi$.

The radiation density formula will have further utility for subsequent discussions in this paper.

Black hole characteristics

Without prejudice to future modifications in black hole modeling, an object within its gravitational or Schwarzschild radius, $r = 2GM/c^2$ will possess black hole characteristics.

The Schwarzschild radius of an object having matter-energy of $1.22 \bullet 10^{10}$ J is about $2 \bullet 10^{-34}$m. A primeval photon occupying an initial spherical locus of Planck diameter will have a radius $8.08 \bullet 10^{-36}$m, which is therefore below its Schwarzschild radius.

It may be worth mentioning here that in virtually all cosmological models [3-5], the universe must at least at some early epoch lie within its Schwarzschild radius. At the Planck era before any possible inflationary scenario, this is especially the case. With $\sim 10^{50}$kg as matter-energy now directly observable, the universe would lie within its Schwarzschild radius when it was of radius $\sim 10^{23}$m, if it had that mass then.

We identify some properties of black holes that may characterize the primeval photon:

- Nothing can classically escape from it. All energy, matter and events in it must remain within its system.
- Black holes can evaporate via the Hawking process [6], if they have an exterior in which particle-antiparticle pairs can be created. However, unlike black holes within an existing universe, the primeval photon has no exterior. All events relevant to it take place within its system. Evaporation via the Hawking process can therefore not apply to it and it could only be extinguished by some other process.
- The Bekenstein-Hawking formulae are used to describe black hole mechanics [7-9]. The equations of interest for a spherically symmetrical black hole are:

$$S_{bh} = A/4 \text{ X } (kc^3/G\hbar) \tag{3}$$

$$A = m^2 \text{ X } 8\pi(G^2/c^4) \tag{4}$$

Substituting Eq.(4) into (3), we also have

$$S_{bh} = m^2 \text{ X } 2\pi(kG/\hbar c) \tag{5}$$

where $\mathbf{S_{bh}}$ is the Bekenstein-Hawking entropy, **A** is surface area of the event horizon of the object, **k** is Boltzmann's constant, ***c*** is light velocity, **G** is the gravitational constant and $\hbar$ is Planck's constant, h divided by $2\pi$, **m** is the mass of the object and **r** its radius.

Leaving the origin of the arrow till later, we see from the equations that if a primeval black hole is born with a thermodynamic arrow imposed on it, its event horizon area (i.e. entropy) must increase till an equilibrium state is attained. More importantly, alongside the increasing entropy, from Eq.(4), its mass or matter-energy must also scale linearly with radius, since $A = 4\pi r^2$.

Objects constantly contained within their Schwarzschild radius obey Eq.(6).

$$r = M \bullet 2G/c^2 \quad (6)$$

If their radius increases, their mass increases, and vice-versa. Since G and $c$ are constants, a linear relationship between the matter-energy and radius can be expected for such objects in any evolution.

Closed system characteristics

"A system created from nothing must be closed and a closed system must have zero net energy and this must be so at ALL epochs"*. This is an alternative statement of the law of conservation of energy. It implies no net energy can be created from 'nothing'. It is a very significant property of closed systems with wide ranging implications for the primeval photon. We will make this our first postulate. The second being that if space is not infinitely divisible into position coordinates, only a single position coordinate will be available at the Planck size ~ $1.616 \bullet 10^{-35}$m.

There is nothing outside of the primeval photon, so any continued evolution while it exists will be adiabatic. From the guiding rules above, though it fluctuates from nothing with an initial energy of $1.22 \bullet 10^{10}$ J ($1.36 \bullet 10^{-7}$kg), this positive matter-energy is counterbalanced by the fact that it has a radius $8.08 \bullet 10^{-36}$m which possesses negative gravitational energy, so the total net energy remains zero. This implies there is no energy cost to the fluctuation. To make things quantitative, a radius of $8.08 \bullet 10^{-36}$m will be equivalent to $- 1.22 \bullet 10^{10}$ J.

Now, if while it is still alive, the radius of the primeval photon's closed system increases from $8.08 \bullet 10^{-36}$m to $10^{25}$m, i.e. the radius increases by a factor of ~ $10^{60}$, this will be a change in negative energy from $- 1.22 \bullet 10^{10}$ J to ~ $-10^{70}$ J.

From the previously identified energy conserving property of closed systems, there must be an accompanying change in the positive matter-energy from $+1.22 \bullet 10^{10}$ J to ~ $+10^{70}$ J to keep total net energy zero. This represents a 'blue-shift' in the initial matter-energy in the system.

It should be noted that this exchange of energy between positive and negative starts right from the fluctuation at time zero. In other words, a positive fluctuation in matter-energy of $+1.22 \bullet 10^{10}$ J must be accompanied by a negative change in gravitational energy or radius of $-1.22 \bullet 10^{10}$ J (in metres, $8.08 \bullet 10^{-36}$m).

If larger changes in radius are subsequently mandated for some reason, we expect any

*See Acknowledgements.

initial positive matter-energy present to be appropriately increased to balance the negative gravitational energy and keep the total energy sum at zero. Such blue-shifted matter-energy will be expected to remain in the system. We depict this in Fig.1.

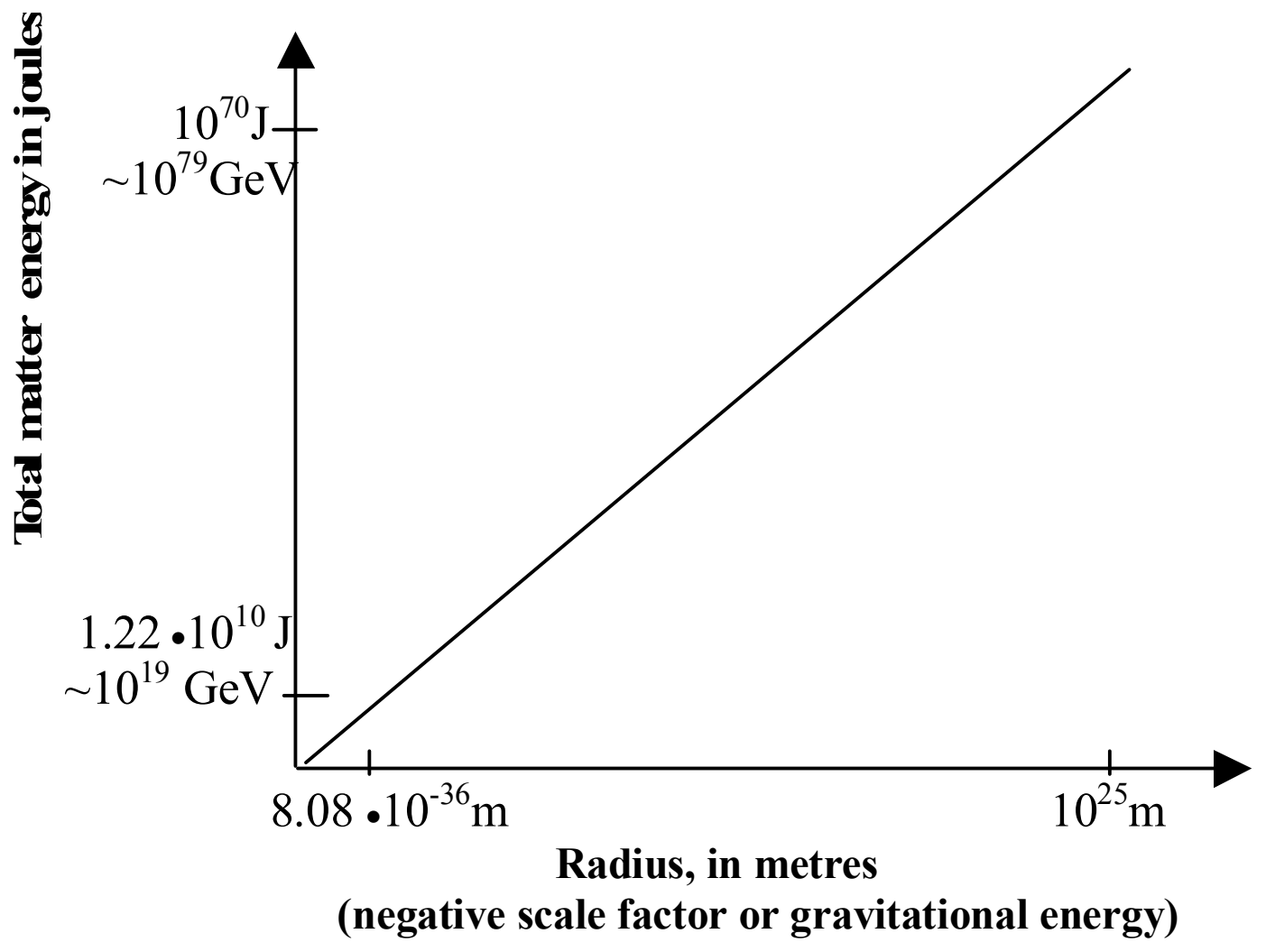

Fig. 1. Showing the implication of energy conservation law for changes in size of a primeval photon's closed system.

The gradient of the slope for the closed system of a primeval photon fluctuating from nothing gives a constant, $\Theta \sim 1.52 \bullet 10^{45}$ $Jm^{-1}$ or in kilogram ($1.68 \bullet 10^{28}$ $kgm^{-1}$). This gives us a quantitative value for what the change in matter-energy would be for the accompanying change in scale factor, i.e. a positive change of $1.52 \bullet 10^{45}$ J in matter-energy for every negative metre change in radius.

Knowing this gradient, $\Theta$, we can quickly estimate the total matter-energy content of the closed system at any era, knowing the radius, R of the system at that era. This is given by

$$\Theta \bullet R \tag{7}$$

At spontaneous fluctuation, it follows from Eq.(7), that when R changes from zero to $8.08 \bullet 10^{-36}$m, the matter-energy content will change from zero to $1.22 \bullet 10^{10}$ J (~$10^{19}$ GeV). When R is zero, matter-energy will be zero. The origin is therefore not an infinitely dense thing occupying zero radius in this scenario, but will be 'nothing'.

In any possible collapsing phase, if the system remains closed and energy conservation still holds, no net positive or negative energy can remain, a reduction in radius will be accompanied by commensurate reductions in positive matter-energy till zero.

It may be noted that all objects along the slope in Fig.1 lie within their Schwarzschild radius and obey Eq.(6). Extrapolating to zero, 'nothing' can rightly lay claim to being an example of a gravitational singularity, though of an extreme and peculiar kind. It obeys $r = 2GM/c^2$ and has Schwarzschild radius of zero. From black hole analogies, its event horizon area (entropy), matter-energy and surface gravity (temperature) are also all zero.

To reiterate, as long as we deal with a closed system whose total energy must remain zero, whether a scale expansion is by an inflationary mechanism or in obedience to some thermodynamic dictate, the positive matter-energy would increase proportionately with the negative scale factor.

## III. Effects

An agency for quantum fluctuation of metric from nothing

A common starting point for inflationary theories of a universe created from nothing is that there must have been a quantum fluctuation of metric, the universe we now behold subsequently arising from the tiny piece of space created.

We have shown above that fluctuation in radius from zero is equivalent to a change in matter-energy from zero as one cannot occur without the other for a closed system. Secondly, the primeval photon must have a wavelength. Thirdly, from dimensional analysis, energy has dimensions of mass, time and space ($kgm^2s^{-2}$). A spontaneous fluctuation of metric is therefore part and parcel of a primeval photon's fluctuation from nothing.

An agency for creation of fundamental matter particles

The conversion of a photon's energy to matter particles, such as an electron and positron is an observed and known phenomenon. No excess charge is however created. The transformation of photons to matter particles would seem most practicable when photon energy is above the gamma frequency ($\sim10^{20}$Hz). This is because an electron has energy $\sim10^{-14}$ J (0.5MeV) and the photon must at least have enough energy to transform into an electron-positron pair. From the energy range $10^{19}$GeV ($10^{10}$J) to $10^{-5}$GeV ($10^{-14}$J), photons will be in the gamma range and will have little difficulty transforming to matter particles. The stability of the matter particle at the prevailing ambient energies is however a different issue. The exact efficiency of conversion of photons to matter may be best further discussed with theories of baryogenesis.

An agency for entropy increase from nothing

The nothing state is a state of perfect order and thermodynamic rest. Entropy, matter, energy, space and time are expected to be absent. Logically, temperature will be absent as well, i.e. absolute zero.

Such a state could be difficult to visualize by observer's who are 'something'. Much in the same way blindness is not easily appreciated by the sighted. For instance, most sighted observer's view blindness as darkness everywhere. But this is incorrect. Darkness itself is still something, much as empty space is. The blind see nothing at all, not even darkness. The reader can try this out by closing one eye and looking through the other. What is seen through the closed eye is not darkness, but 'nothing'. The eye seems not to exist any longer. Nothing is an absence of all that characterizes existence, including spatial extent and positions. There can be no disorder or disequilibrium in such a state. It must be a state of perfect order.

The second law of thermodynamics tells us what happens to entropy when there is a change in the matter-energy of a state at equilibrium. This is governed by Clausius' equation

$$\partial S = \partial E / T \tag{8}$$

where $\partial E$ is the change in matter-energy, T is the temperature at the moment matter-energy change takes place and $\partial S$ is the effected change in entropy, ( i.e. change in the number of position and momentum coordinates available to the system for arrangement in different possible ways). As earlier mentioned, it is not the net energy (which can remain zero), but the amount of positive matter-energy present that determines the thermodynamic properties of a closed system.

Eq.(8) above, may be viewed as incomplete without adding the work terms, $P\partial V$. We leave these out because of the conceptual difficulties associated with whether a system which is 'all there is' can do work on itself or on a non-existent exterior. However, even if included any resulting entropy change would only be further increased, since the work term will be positive when a system increases in size.

According to Eq. (8), if the fluctuation of a primeval photon causes a positive change in the matter-energy of a state at absolute zero, there will be an astronomical increase in the entropy of the system and the photon's energy must become spread over more degrees of freedom. If space is not infinitely divisible into position coordinates and only one position coordinate is available at the Planck length, there will have to be further increases in system size to accommodate the physically and mathematically mandated additional position coordinates.

Experience in thermodynamics suggests that effected increase in entropy is not instantaneous but occurs gradually and continuously till the equilibrium value of entropy dictated by Eq.(8) is attained. If equilibrium entropy value lies towards infinity for a matter-energy change at absolute zero, the initial conditions present at the time of the primeval photon's fluctuation would seem to bequeath it with a thermodynamic arrow pointing towards increasingly higher entropy values despite its evolution being adiabatic and despite remaining a closed system.

The spreading of the photon's energy over more degrees of freedom represents a 'red-shifting' in energy and thus lower frequency. This is in contrast to the earlier mentioned

'blue-shifting' that accompanies increase in radius and which is required to keep total energy sum zero.

In black hole thermodynamics, the analogue of Clausius equation, Eq.(8) is given by

$$\partial A = \partial E / (\kappa/2\pi) \qquad (9)$$

leaving out terms like charge and angular momentum [6-9]. $\kappa$ is the surface gravity at the time of change in matter-energy, $\partial E$. In black hole mechanics, $\kappa/2\pi$ is analogous to temperature, while change in event horizon area, $\partial A$ is analogous to entropy change, $\partial S$.

For a beginning from an infinitely dense thing, the initial surface gravity and thus temperature would be astronomical. This has been one way of looking at the big bang scenario.

As earlier suggested and as can be extrapolated to the infinitesimal from Fig.1, if nothing represents a gravitational singularity of zero matter-energy and thus lying within zero Schwarzschild radius and zero event horizon area, its initially surface gravity, i.e. temperature will also be zero. What will be the thermodynamic effect of a quantum fluctuation in matter-energy on such a state?

The equations of black hole mechanics seem to suggest that there will be an increase in the event horizon area. Furthermore, Eq.(9) tells us by how much the event horizon area might be expected to change.

We therefore expect that a black hole born in these circumstances would harbor a thermodynamic arrow. For black holes both the thermodynamic arrow and increase in size ( ?cosmological arrow) lie in the same forward direction of time.

**IV. Life History and Life Expectancy**

<u>Life Expectancy</u>

Our current theories wish the primeval photon dead within about $10^{-44}$ seconds. Among the theoretical grounds for this wish are the uncertainty principle, the Hawking evaporation process and the possible gravitational collapse to infinite density.

If in spite of these threats, the primeval photon is still alive and is the same photon smiling at us through the cosmic background radiation, CBR, how could it have survived? We look at the possibilities, some of which may appear contrived but we make no excuses for this because we are looking for possible escape routes in case the primeval photon could possibly still exist.

The Hawking evaporation process has been taken care of by the fact that the primeval photon has no exterior in which particle-antiparticle pairs can be created unlike black holes in an existing universe.

From gravitational theory alone, since the primeval photon is born within its Schwarzschild radius, the matter-energy should become concentrated gradually to zero radius and not have its energy distributed over more degrees of freedom. However, we have seen that having fluctuated from nothing, the primeval photon is a closed system whose total net energy must sum to zero. If the radius collapses to zero and the matter-energy retains a positive value, the energy conservation property of closed systems will be violated. Its positive matter-energy must change with its negative radius to keep its net energy zero. Gravitational expectations alone may however not be the only theory governing its existence. There is an existing thermodynamic mandate for the primeval photon's energy to become spread over more position and momentum coordinates rather than collapse to a single position coordinate in phase space. The contending theories may however find a middle ground, retaining the system within its gravitational radius on one hand, whilst on the other allowing increase in the number of position coordinates available in it for the different possible arrangements that define entropy increase.

The uncertainty relation is the more potent threat [10] and we examine this threat by asking questions concerning its scope of validity. The original interpretation of the relation has to do with measurement accuracy of conjugate variables and not their absolute magnitude. For instance, for position and momentum we deal with their precision of measurement. If we can determine position with absolute certainty, the momentum becomes absolutely uncertain to the extent of the relation

$$\Delta p \bullet \Delta m \geq h/2\pi \qquad (10)$$

The differential sign is the error in measurement and not the absolute magnitude of the conjugate parameter. For energy and time, the uncertainty relation is

$$\Delta E \bullet \Delta t \leq h/2\pi \qquad (11)$$

Since the original premise for the uncertainty relation is as above, could the energy-time relation as applicable to the primeval photon have to do with the precision in measurement also and not the absolute magnitude of the conjugate variables, such as the duration of existence? If its wavelength and thus energy is known with certainty can its life expectancy also be measured with certainty? Will measuring its duration with certainty contravene the original premise for formulating the uncertainty principle?

Again, since no net energy is actually created and energy conservation law is not violated at any time, will the energy-time uncertainty relation still apply to the primeval photon?

Another means of escaping the death threat from interpretations of the uncertainty relation is that the relation makes allowance for 'real' and 'virtual' fluctuations. Virtual particles being those that cannot exceed their pertinent survival times, while real particles are those that do. Although virtual particles have never been seen, their effects are described experimentally. The identification of a particle as real or virtual is from experience. A particle being labeled virtual if it does not outlive its pertinent lifetime and labeled real if it does. If the primeval photon is a real particle, then it must still exist and have something to do with our observed universe.

A third way out is to take a reverse perspective. That is, instead of viewing the primeval photon as an initial fluctuation in matter-energy, we can equivalently view it as a fluctuation of metric or radius, which it equally is. In this view, time would not be the conjugate variable determining its lifespan.

Aside the above, others may find other compelling thermodynamic or gravitationally inspired mechanisms, e.g. within inflation theory, that may intervene to preserve the primeval photon. These life-saving devices are in our opinion worth further study.

Thermal history

We now review the thermal history of the primeval photon assuming it is not extinguished prematurely as theoretically envisaged.

During the system's history, no net energy is created at any era, even at the time of initial fluctuation as earlier demonstrated. Positive matter-energy always tracks and balances the negative gravitational energy of the radius. However, as volume changes as cube of the radius, R, matter-energy density falls as $R^{-3}$. These changes have positive and negative implications for temperature and ambient energy. Positive implications because matter-energy in the system increases linearly with radius, and negative because despite this, the matter-energy density still falls with increasing radius.

We can know the total matter-energy at any epoch from Eq.(7), $\Theta \bullet R$, where R is the radius at that epoch.

For eras, when most of the positive matter-energy will be in form of photons, i.e. during radiation-dominated eras, the radiation density formula, $E_D = aT^4$, applies and will be useful to calculate the thermal history of the primeval photon's system.

Energy density, $E_D$ at an epoch is total matter-energy at that epoch divided by the volume at the epoch

$$\therefore \quad E_D = \Theta \bullet R / (4\pi R^3)/3 \qquad (12)$$

At radiation-dominated epoch's, $E_D = aT^4$. Therefore in those era

$$aT^4 = \Theta \bullet R / (4\pi R^3)/3 \qquad (13)$$

Rearranging and canceling out, we have

$$T^2 = 1/R \bullet \sqrt{3\Theta/(4\pi a)} \qquad (14)$$

All the entities in square root are constants for a primeval photon fluctuating from nothing and together have a value ~ $6.88 \bullet 10^{29}$ $K^2m$

$$\therefore \quad T^2 = 1/R \bullet (6.88 \bullet 10^{29}) \qquad (15)$$

Eq.(15) will be accurate during radiation-dominated eras and can be used to quickly estimate the thermal history of a system arising from a primeval photon fluctuation during those eras. Significantly, it incorporates the changes in matter-energy due to increase in scale factor (the blue-shifting) as well as reductions in energy density due to same change (the red-shifting).

For latter eras, when primeval energies may have partially transformed to matter and possibly other non-radiating forms of energy, the accuracy is expected to diminish for the calculation of temperature.

Using Eq.(15), the following temperatures are obtainable quickly, as well as the corresponding ambient energies from E ~ kT. (1 GeV ~ $10^{13}$ K)

| Time (seconds) | Radius (metres) | Temperature (kelvin) | Ambient matter-energy (GeV) |
|---|---|---|---|
| 0 | 0 | 0 | 0 |
| $10^{-44}$s | $8.08 \bullet 10^{-36}$m | $2.9 \bullet 10^{32}$K | $10^{19}$GeV |
| $5.4 \bullet 10^{-44}$ | $1.616 \bullet 10^{-35}$ | $2.07 \bullet 10^{32}$ | $10^{19}$ |
| $10^{-35}$ | $3 \bullet 10^{-27}$ | $10^{28}$ | $10^{15}$ |
| $10^{-33}$ | $3 \bullet 10^{-25}$ | $10^{27}$ | $10^{14}$ |
| $10^{-30}$ | $3 \bullet 10^{-22}$ | $10^{25}$ | $10^{12}$ |
| $10^{-10}$ | 0.1 | $10^{15}$ | $10^{2}$ |
| $10^{-6}$ | 300 | $10^{13}$ | 10 |
| $10^{-3}$ | 300,000 | $10^{12}$ | $10^{-1}$ |
| 100 | $3 \bullet 10^{10}$ | $10^{9}$ | $10^{-4}$ |
| $10^{13}$ | $3 \bullet 10^{21}$ | $10^{4}$ | $10^{-9}$ |

Table 1. Showing thermal history of a long-lived primeval photon's universe.

*There is presumably no length smaller than the Planck length so the second line in Table 1 refers to the Planck diameter of the primeval system.

The table is extrapolated to cover cosmologically significant or speculated epochs in the standard big bang model, such as the 'inflationary epoch' (before $10^{-32}$s), 'desert epoch' (~$10^{-32}$ to $10^{-10}$s) and the recombination or photon decoupling epochs (3 minutes to about 300,000 years).

The thermal history of the standard big bang model [3-5] is principally derived from the ambient energies and temperatures that guarantee the stability of the fundamental forces and structures at each epoch. While that of the primeval photon is derived from what we know of photons and the characteristics of closed systems. It is remarkable that despite this independent derivation there is a general agreement of the thermal history of the primeval photon with that of the standard big bang model except for the following contentious areas:

(i) The primeval photon's history is linear and does not seem to require a period of exponential change in radius before there is agreement with the temperatures and ambient energies of the standard big bang model.

(ii) The relationship between temperature and radius is more accurately $T^2 \bullet R \sim$ constant, and not $T \bullet R \sim$ constant during radiation-dominated era.

(iii) Inflation theory [11,12] describes pumping of energy from the exponentially increasing radius to the matter-energy fields during the inflationary epoch. If the system remains closed after the epoch and energy conservation law still holds, the further linear increases in radius (negative energy) must still be compensated by increases in matter-energy (positive energy), so that the total energy sum remains zero and energy is not created or destroyed in the post-inflation era. Current inflationary literature does not seem to discuss this to best of our knowledge.

(iv) At $10^{-10}$ seconds, the thermal history of the primeval photon predicts a temperature of $10^{15}$ K, when the system was about 10cm in size, with ambient energy ~ $10^2$ GeV. This agrees well with what is predicted by the standard big bang model for a universe of that life span so that the Salam-Weinberg phase transitions separating the electroweak into electromagnetic and weak nuclear forces can take place.

However if there is an inflationary epoch it will take place and end anytime before $10^{-32}$ seconds from the beginning. During such an epoch, depending on the inflation model temperature drops rapidly to ~ $10^{22}$ K before increasing rapidly and ending up at ~ $10^{27}$ K (i.e. $10^{14}$GeV) with the radius equally increasing exponentially from about $10^{-27}$ m to 0.1m. This causes a severe 'temperature problem' for the standard big bang model. From the radiation density formula, for our observable universe of 0.1m radius to have a temperature $10^{27}$ K after inflation, about $10^{90}$J of energy must have been produced and contained in it. This enormous amount of energy would lead to inconsistent temperatures and ambient energies throughout the radiation-dominated era from 0.1m to ~$10^{21}$m. To be precise, inflation scenario with this energy content will give us $10^{27}$K ($10^{14}$ GeV) at 0.1m ($10^{-10}$s); $10^{24}$K at 300m ($10^{-6}$s); $10^{18}$K at $3 \bullet 10^{10}$m (100s) and $10^9$K at $10^{21}$m radius ($10^{13}$ seconds). This will be grossly in conflict with the thermal history of the standard big bang model for the radiation-dominated era. At $10^{-10}$s, the physics becomes less speculative and the problem of how to reach an acceptable model temperature of $10^{15}$ K (~$10^2$GeV) at that time given the applicable energy densities at the end of an inflationary epoch is problematic. Energy dissipating processes will become necessary to preserve correspondence with the standard big bang model. Such processes may be theorized as possible during the 'desert' epoch after inflation, but the physics will be highly speculative.

(v) The matter-energy directly observable in the universe now with radius ~ $10^{25}$m is above $10^{50}$ kg (i.e. $10^{67}$J). This is close to but not up to the critical density. The standard big bang model does not specifically tell us how and when the universe acquired this amount of matter-energy. Has it been there from the beginning or was it after an inflationary epoch? This quantity of matter-energy gives discrepancies in the temperature of the universe as depicted in the standard big bang model if it was there during the radiation-dominated era. Again using the radiation density formula, the universe would have had temperature ~$10^5$K instead of less than $10^4$K at radius $10^{21}$m and temperature $10^{21}$K instead of $10^{15}$K at radius about 0.1m as favored by the big bang model. If it had been there at the Planck time, temperature would have been above $10^{47}$K ($10^{34}$GeV), instead of ~ $10^{32}$K ($10^{19}$ GeV). If dark matter-energy is added, then the universe would be even hotter during those radiation-dominated eras. It is therefore obvious that the amount of energy in those eras would have to be far less than the amount of matter-energy observable in the universe today for the standard big bang model to remain consistent. A continuous increase in the quantity of matter-energy harbored in the universe with increase in its radius however brings a consistency to the thermal history of the universe as depicted by the standard big bang model. From the Planck era up to the end of the radiation-dominated eras, this is demonstrable from the equations above.

(vi) The photon decoupling or recombination era is speculated to last from about 3 minutes to 300,000 years after the beginning. That is from radius $3{\bullet}10^{10}$m to $10^{21}$m or temperatures $10^9$K to ~ 3000K. If the CBR is a mixture of photons decoupled at the widely varying temperatures of this era, the homogeneity in the CBR may probably not be as much as that observed. The initial primeval photon, being an actual black body may be a source for the CBR, if long-lived. The homogeneity of the photons in the CBR may also come naturally, if they are relics of it.

We end the primeval photon's history with a couple of unanswered questions. If temperature, like space, matter and energy cannot be infinitely small, how long would it take the system to cool to absolute zero? Is there a largest possible black hole and what will be its size? Would its surface gravity, and thus temperature, be about absolute zero? What will be the consequences of this? Literally speaking, can a hole ever be filled up or does it get infinitely deeper? Would matter-energy and entropy be zero in an absolute zero state? If matter-energy would be zero, from the energy conservation laws we have previously highlighted, would this be equivalent to zero radius and would we arrive at a nothing state again? We leave further speculations about the primeval photon's end for the future.

## V. Concluding Remarks

Given the known properties of closed systems, the characteristics, effects and behavior of a primeval photon over a scale ranging from Planck size to ~ $10^{21}$m appears to accord well with what is described in the standard big bang model, particularly the temperatures and ambient energies applicable at each epoch.

If space is not infinitely divisible into positions, then only a single particle must have occupied the only available position coordinate at the Planck size. Historically, Georges Lemaitre, a Belgian priest and mathematician had anticipated this point and used it as a working hypothesis [13]. An atom, electron or other matter particle as we know them would however be too big at that time. Their de Broglie wavelength would also not fit. Only a photon of that wavelength would fit. The temperature and ambient energy of such a photon corresponds remarkably with what is predicted by the standard big bang model for that epoch, as depicted in Table 1.

The emergence of such a primeval photon from nothing as well as any subsequent evolution seems to preserve the law of energy conservation at all epochs. The total energy of the system always summing to zero, which is the same as the energy of the 'nothing' state.

Since it does not require any net positive energy to fluctuate or change its radius, even from zero, the system does no work nor is any work done on it. It may therefore not be a caused phenomenon but a spontaneous one.

The motivation for change in size appears located in the initial thermodynamic state before fluctuation. If a mathematical and physical necessity for entropy increase is imposed on the system, then additional position coordinates that will make more ways of arrangement available must subsequently emerge in the closed system. Such increases in size and radius will however remain consistent with the closed system characteristics we have identified causing a 'blue-shift' of contained matter-energy to balance the negative energy of increasing radius. The 'blue-shifting' of energy with R is however in contrast to the 'red-shifting' that is due to reducing energy density which scales as $R^{-3}$. These contrasting factors are incorporated in the equations used to calculate the thermal history of a long-lived primeval photon.

Other inferences derivable from the closed system characteristic of net zero energy include the fact that if mass and radius are not constant but increase, both must do so linearly with time. For a universe, now $\sim 10^{25}$m, it must have a mass now $\sim 10^{53}$kg. For a long-lived primeval photon, the ratio of mass to radius is about $1.52 \bullet 10^{45}$ $Jm^{-1}$ (or $1.68 \bullet 10^{28} kgm^{-1}$) and is constant in time.

For other objects evolving within their gravitational radius, a similar ratio of mass to radius will apply, since this ratio can also be derived from $c^2/2G$, but it may have a slightly lower value than that of the primeval photon which lies an additional two factors of ten below its Schwarzschild radius.

The positive matter-energy required to balance the negative radius must be distributed among what is available to store it. We have seen that with reducing ambient energies, photons energetic enough to transform to matter become less. The positive matter-energy initially conveniently stored as photons, then later as a mix of photons and matter may as new matter and photons become increasingly unavailable after $10^{13}$ seconds of expansion become stored in non-matter and non-radiation forms. As is commonly speculated, we therefore envisage an initial photon-dominated era, followed by a matter-dominated one and ending up with a dark matter-energy dominated era. The initial matter-energy density $\sim 10^{114}$ $Jm^{-3}$ will now at $\sim 10^{25}$m become about $10^{-5}$ $Jm^{-3}$, a dilution factor $10^{120}$ times lower than the earliest epoch.

If the primeval photon we wish dead within $\sim 10^{-43}$ seconds is still the same one smiling at us through the CBR, perhaps we need to take it serious. For one, it has no horizon, homogeneity or isotropy problems since the photons in the CBR will in the main be relics of a single primeval photon. Secondly, it may not be plagued by monopole problems, since none is described. Thirdly, growth in scale factor seems not to require an exponential rate, a linear rate at velocity $c$ agreeing fairly well with the ambient energy and thermal history depicted in the standard big bang model, particularly during the radiation-dominated era when the radiation density equation is most useful. Fourthly, it does not have a singularity problem since there is no need for matter-energy to occupy a zero radius. Since it remains within its gravitational radius at all eras, no matter any changes in scale, it also does not have a flatness problem at any era. The positive matter-energy required to keep the system closed co-evolves with the negative gravitational energy of the radius. This is a reflection of obedience to the energy conservation law and an indicator of the zero energy state of the origin. The parameter, $\Omega$ which is the ratio of the amount of matter-energy present to the amount required to keep the system closed and within its Schwarzschild radius therefore remains unity at all epochs despite over $10^{60}$ Planck times of expansion. At the Planck era, the Planck mass is required to keep the system closed. Now the amount of matter-energy required may be about $10^{53}$kg, if radius is $\sim 10^{25}$m. Finally, it does not have a 'temperature problem', a problem not commonly discussed but which afflicts the standard big bang model and any inflationary modifications.

If the primeval photon brings increased consistency to the standard big bang model, perhaps we need to reconsider its theoretical death sentence.

**Acknowledgements**

I thank Alex Vilenkin for stubbornly insisting that "a universe created from nothing must be closed and that a closed system must have zero net energy at all epochs". This was instrumental in bringing an overall consistency to the thermal history of the primeval photon's universe as presented here.

## References

1. Tryon, E.P., Is the universe a vacuum fluctuation? *Nature*, **246**, 396, (1973).

2. Vilenkin, A., Creation of universes from nothing. *Phys. Lett.*, **117B**, 25, (1982).

3. Kaufmann, W.J. III, *Universe*. (Freeman, New York, 1987).

4. Silk, J., *The Big Bang*. (Freeman, New York, 1989).

5. Weinberg, S. *The First Three Minutes*. (Basic Books, New York, 1988).

6. Hawking, S.W. Particle creation by black holes. *Comun. Math. Phys*., **43**, 199-220, (1975).

7. Bekenstein, J.D., Black holes and entropy. *Phys. Rev*. D, **7**, 2333-46, (1972).

8. Bardeen, J.M., Carter, B. and Hawking, S.W., The four laws of black hole mechanics. *Commun. Math. Phys*., **31**, 161-70, (1973).

9. Bekenstein, J.D., Black hole thermodynamics. *Physics Today*, January, (1980).

10. Polkinghorne, J.C., *The Quantum World*. (Princeton University Press, Princeton, N.J., 1984).

11. Linde, A.D., Inflation and quantum cosmology. In *300 Years of Gravitation*, Eds. S.W. Hawking and W. Israel, (Cambridge University Press, Cambridge, 1987).

12. Blau, S.K. and Guth, A.H., Inflationary cosmology. In *300 Years of Gravitation*, Eds. S.W. Hawking and W. Israel, (Cambridge University Press, Cambridge, 1987).

13. Lemaitre, G. *The Primeval Atom*. (D. Van Nostrand, New York, 1950).